\documentclass{article}
\usepackage{frascatiphys_R}
\input{epsf}
\def\mevc {\ifmmode {\rm MeV}/c \else MeV$/c$\fi}
\def\mevcc {\ifmmode {\rm MeV}/c^2 \else MeV$/c^2$\fi}
\def\gevc {\ifmmode {\rm GeV}/c \else GeV$/c$\fi}
\def\gevcc {\ifmmode {\rm GeV}/c^2 \else GeV$/c^2$\fi}
\def\ra   {\rightarrow}
\newcommand{\Bs} {\ifmmode B_{\mbox{\sl s}}^{0}
                       \else $B_{\mbox{\sl s}}^{0}$\fi}
\newcommand{\Ds} {\ifmmode D_{\mbox{\sl s}}^{+}
                       \else $D_{\mbox{\sl s}}^{+}$\fi}
\newcommand{\dms} {\ifmmode \Delta m_{\mbox{\sl s}} \else 
                           $\Delta m_{\mbox{\sl s}}$\fi}
\begin{document}
\title{ 
Results on Heavy Quark Physics at TeV Energies
}
\author{
Manfred Paulini 
\thanks{\hspace{0.2cm} Representing the CDF and D\O\ Collaboration.}\\
\em Carnegie Mellon University, Pittsburgh, PA 15213, U.S.A.}
\maketitle
\baselineskip=11.6pt
\begin{abstract}
We review recent result on heavy quark physics at TeV energies focusing on
Run\,II measurements from the CDF and D\O\ experiments at the
Tevatron.
%Run\,II measurements from the CDF and D\O\ experiments operating at the
%Fermilab Tevatron Collider.
\end{abstract}
\baselineskip=14pt
\section{Introduction}
The CDF and D\O~experiments can look back to an already
successful heavy flavour physics program during 
the 1992-1996 Run\,I data taking period (for a review of $B$~physics
results from CDF in Run\,I see e.g.~Ref.\cite{ref:myrevart}).
The Fermilab accelerator complex has undergone a major upgrade in
preparation for Tevatron Run\,II.
The centre-of-mass energy has been increased from 1.8~TeV to 
1.96~TeV and the Main Injector, a new 150~GeV
proton storage ring, has replaced the Main Ring as injector of protons and
anti-protons into the Tevatron.

The initial Tevatron luminosity steadily increased throughout Run\,II.
By the summer of 2004, the peak luminosity reached 
is $\sim\!10\times 10^{31}$~cm$^{-2}$s$^{-1}$.
The total integrated luminosity delivered by the Tevatron to CDF and
D\O\ by the time of this conference 
is $\sim\!400$~pb$^{-1}$.
More than 300~pb$^{-1}$ were recorded to tape by each CDF and D\O.
However, most results
shown in this review use about 150-250~pb$^{-1}$ of data.
The CDF and D\O\ detectors have also undergone major upgrades
for Run\,II which can be found
elsewhere\cite{ref:docdfup}. 

\subsection{Triggering on Heavy Quark Decays}

The total inelastic $p\bar p$ cross section at the Tevatron is 
about three orders of magnitude larger than the $b$~production cross
section. The CDF and D\O\ trigger system is therefore the most
important tool for finding $B$~decay products. 
$B$~physics triggers at CDF and D\O\ are based on
leptons including single and dilepton triggers. 
Identification of
dimuon events down to very low momentum is possible,
allowing for efficient $J/\psi \rightarrow \mu^+\mu^-$ triggers. 
Both experiments also use inclusive lepton triggers designed
to accept semileptonic $B\rightarrow \ell \nu_\ell X$ decays.
New to the CDF detector is the ability to select events
based upon track impact parameter.  
The Silicon Vertex Trigger (SVT)
gives CDF access to purely hadronic $B$~decays and makes CDF's $B$~physics
program fully competitive with the one at the
$e^+e^-$~$B$~factories. 

\section{Selected Heavy Quark Physics Results from the Tevatron}

With the different $B$~trigger strategies above, the
Collider experiments are able to trigger and reconstruct large samples
of heavy flavour hadrons. 
Due to the restricted page limit for these
proceedings, we can only very briefly discuss a few selected
heavy quark physics results from CDF and D\O\ in the following. 

\subsection{$B$~Hadron Masses and Lifetimes}

\begin{figure}[tb]
\centerline{
\epsfxsize=5.8cm
\epsffile{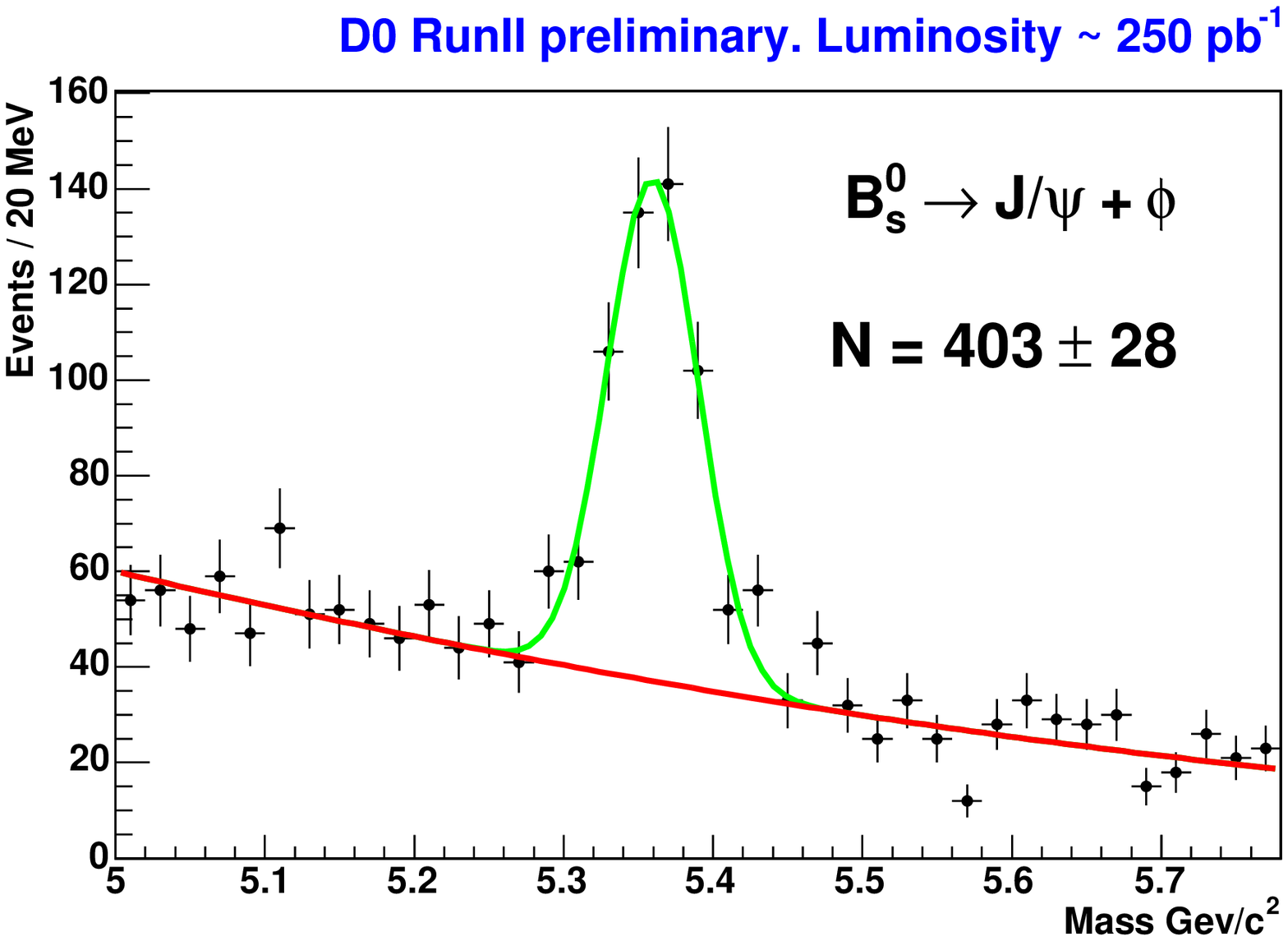}
\epsfxsize=6.2cm
\epsffile{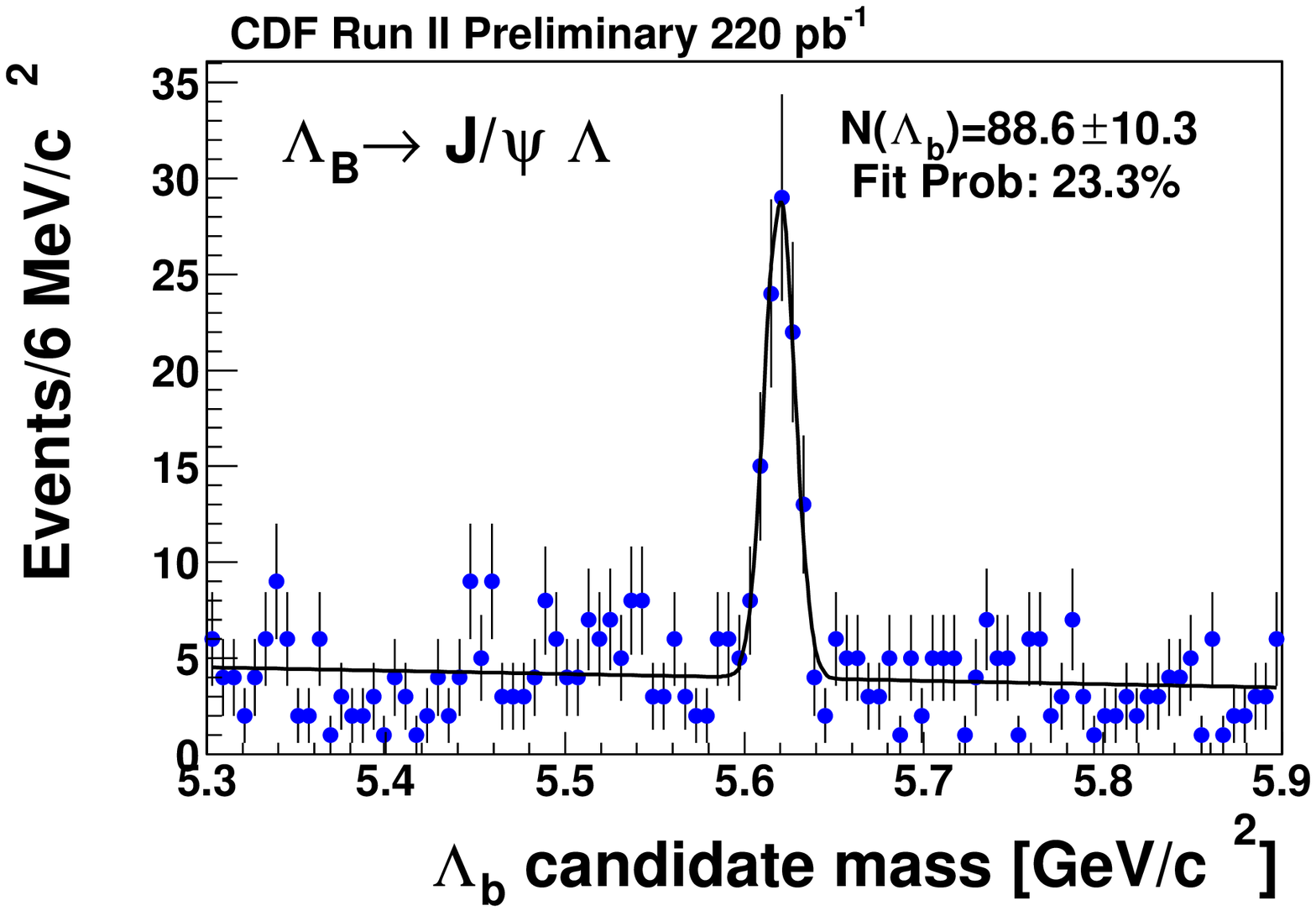}
\put(-320,90){\bf (a)}
\put(-135,80){\bf (b)}
}
\caption{\it
Invariant mass distribution of (a) $J/\psi\phi$ (D\O) and (b) 
$J/\psi\Lambda$ (CDF).
}
\label{fig:bmasses}
\end{figure}

Measurements of $B$~hadron masses and lifetimes are basic calibration
measures to demonstrate the understanding of heavy flavour
reconstruction. CDF and D\O\ use exclusive $B$~decay modes into
$J/\psi$ mesons for precision measurements of $B$~hadron masses
reconstructing the decay modes 
$B^0\ra J/\psi K^{\ast 0}$,
$B^+\ra J/\psi K^+$, $\Bs\ra J/\psi\phi$ and 
$\Lambda_b\ra J/\psi\Lambda$ (see Fig.~\ref{fig:bmasses}).
These modes combine good signal  
statistics with little background.  
The results of the mass and corresponding $B$~hadron lifetime measurements
are summarized in 
Table~\ref{tab:bmasslife}. The \Bs\ and $\Lambda_b$ masses and lifetimes
are currently the world best results.

%-------------------------------------------------------------------
%--------------------------------------------------------------------

%///////////////////////////////////////////////////////////////////////
\begin{table}[tbp]
\centering
\caption{\it
Summary of $B$~hadron mass $m_B$ and lifetime $\tau_B$ measurements from
CDF and D\O. 
}
\vskip 0.1 in
\begin{tabular}{cccc}
\hline
Mode & $m_B$ (CDF) & $\tau_B$ (CDF) & $\tau_B$  (D\O) \\
 & [\mevcc] & [ps] & [ps] \\
\hline
$B^0\ra J/\psi K^{\ast 0}$ & $5279.6\pm0.5\pm0.3$ & 
$1.54\pm0.05\pm0.01$ & $1.47\pm0.05\pm0.02$ \\
$B^+\ra J/\psi K^+$  & $5279.1\pm0.4\pm0.4$ & 
$1.66\pm0.03\pm0.01$ & $1.65\pm0.08^{+0.09}_{-0.12}$ \\
$\Bs\ra J/\psi\phi$ & $5366.0\pm0.7\pm0.3$ & 
$1.37\pm0.10\pm0.01$ & $1.44\pm0.10\pm0.02$ \\
$\Lambda_b\ra J/\psi\Lambda$ & $5619.7\pm1.2\pm1.2$ & 
$1.25\pm0.26\pm0.10$ & $1.22\pm^{+0.22}_{-0.18}\pm0.04$ \\
\hline
\end{tabular}
\label{tab:bmasslife}
\end{table}
%///////////////////////////////////////////////////////////////////////

\subsection{Measurement of Lifetime Ratio $\tau(B^+)/\tau(B^0)$}

\begin{figure}[tb]
\centerline{
\epsfxsize=6.0cm
\epsffile{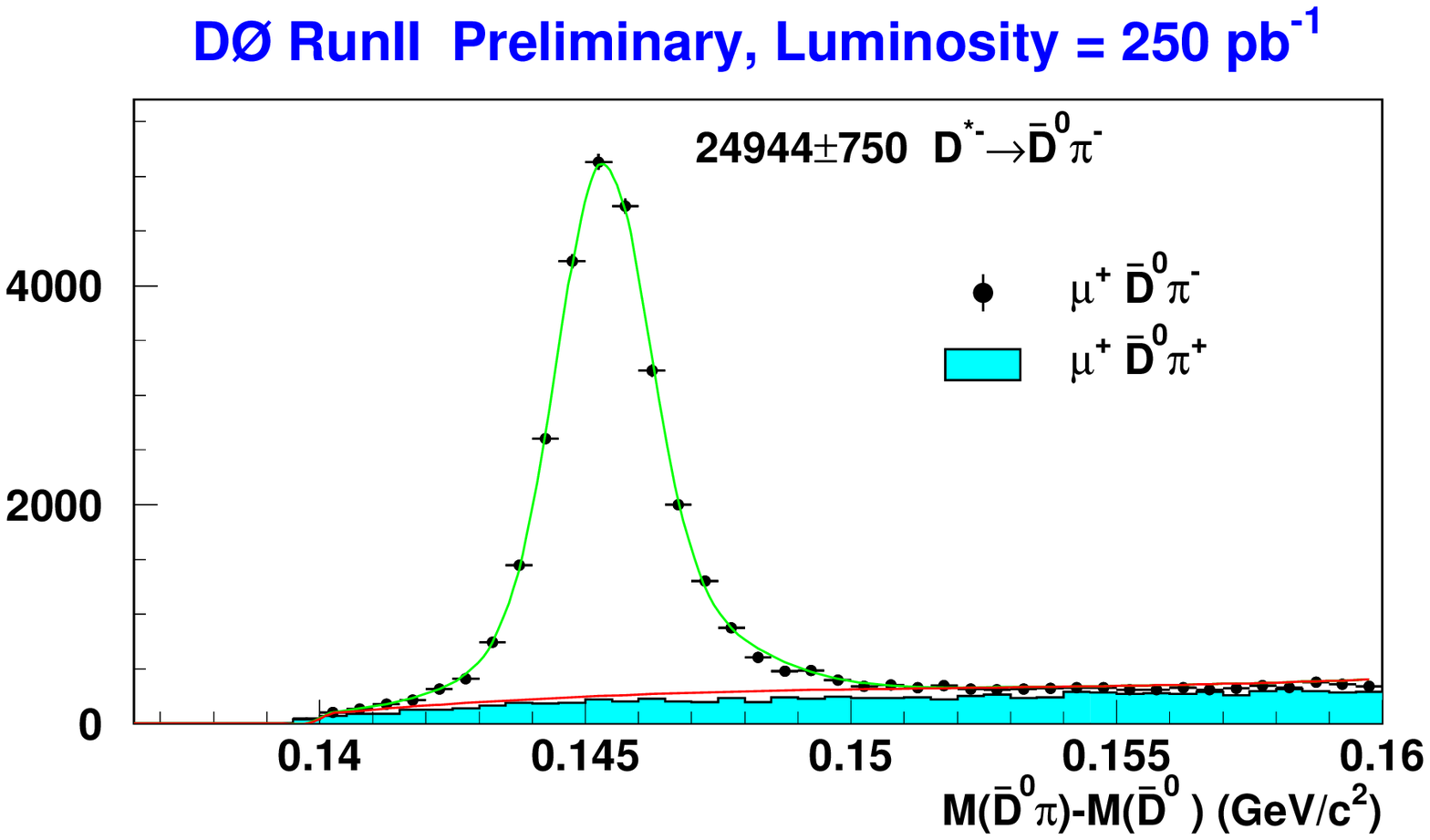}
\epsfxsize=6.0cm
\epsffile{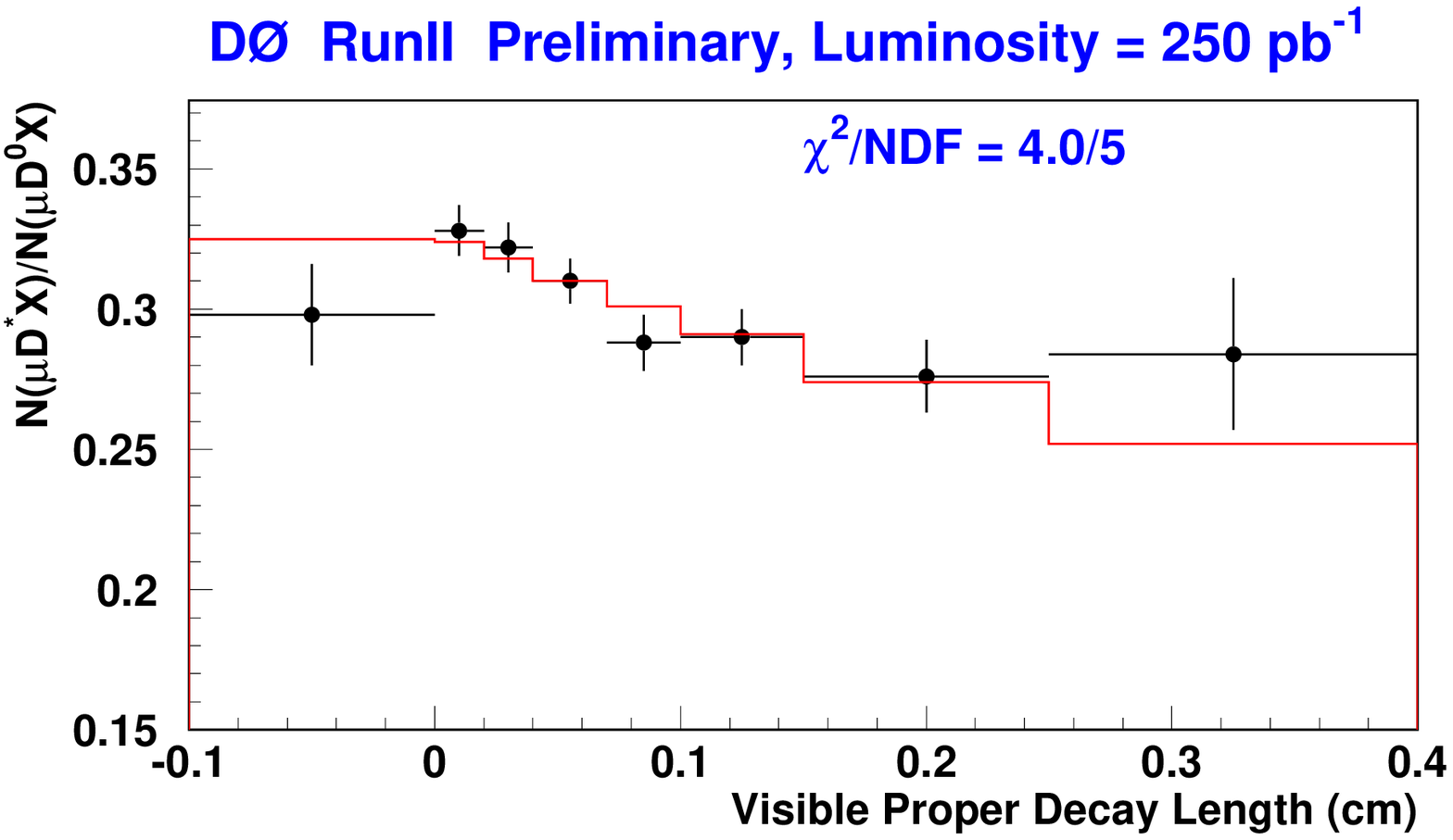}
\put(-320,70){\bf (a)}
\put(-140,30){\bf (b)}
}
\caption{\it
(a) Mass difference $m(\bar D^0\pi)-m(\bar D^0)$ for $\bar D^0\mu$
  events. 
  (b) Ratio of events in $D^{*-}\ (B^0)$ and 
  $\bar D^0\ (B^+)$ samples as a function of
  proper decay length.
}
\label{fig:blife}
\end{figure}

The study of heavy flavour lifetimes is intimately related with the
understanding of the decay dynamics of these particles. The D\O\ experiment
measured the lifetime ratio for neutral and charged $B$~mesons using a
novel technique. This result exploits the large semileptonic sample of
$B\ra\mu X$ decays
reconstructed in about 250~pb$^{-1}$ of $p\bar p$~data. Rather than
measuring the individual $B^0$ and $B^+$ lifetimes and forming the ratio,
this analysis makes use of the fact that $D^{*-}\mu^+$ events mainly
originate from $B^0$~mesons ($\sim 86\%$) while $\bar D^0\mu^+$ indicate a
$B^+$ signature ($\sim 82\%$). The construction of the $B$~decay vertex
uses only $\bar D^0\mu^-$ while the slow pion from the decay 
$D^{*-}\ra \bar D^0 \pi^-$ is only used to distinguish $B^0$ from $B^+$
(see Fig.~\ref{fig:blife}(a))
drastically reducing the systematic uncertainty between both decay modes.
The events are grouped into bins of proper decay length and the $\bar D^0$
event 
yield is extracted from the $K^+\pi^-$ mass distribution. 
Feed-down from $D^{**}$ decays is accounted for
using Monte Carlo studies. The ratio of events in the $D^{*-}\ (B^0)$ and 
$\bar D^0\ (B^+)$ samples as a function of proper decay length, as shown in 
Fig.~\ref{fig:blife}(b), is used to extract a lifetime ratio of 
$\tau^+/\tau^0=1.093\pm0.021\pm0.022$. This is one of the most precise
measurements of the $B^+/B^0$~lifetime ratio.

%-------------------------------------------------------------------
%--------------------------------------------------------------------

\subsection{Charmless $B$~Decays}

\begin{figure}[tb]
\centerline{
\epsfysize=4.7cm
\epsffile{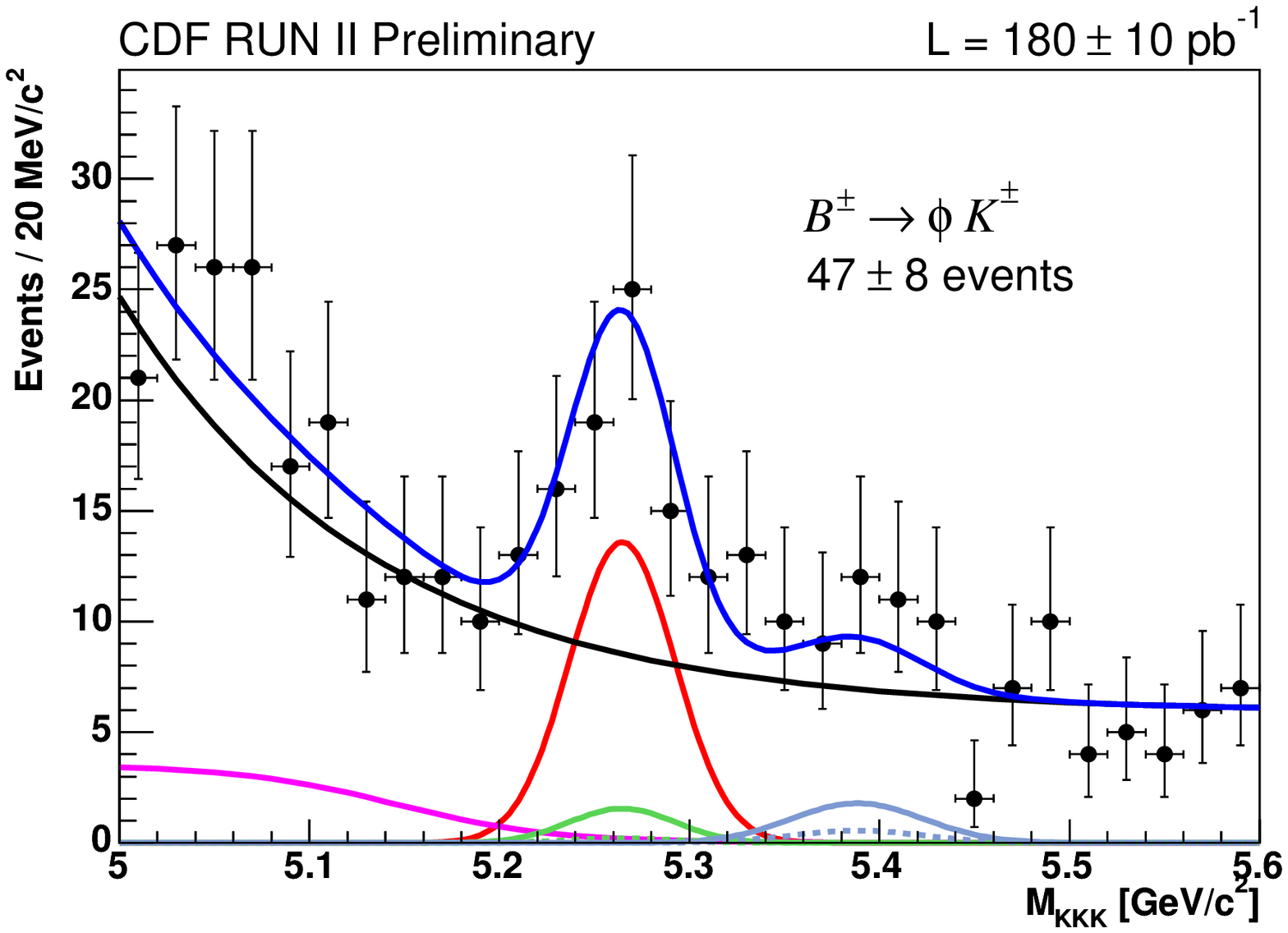}
\epsfysize=4.7cm
\epsffile{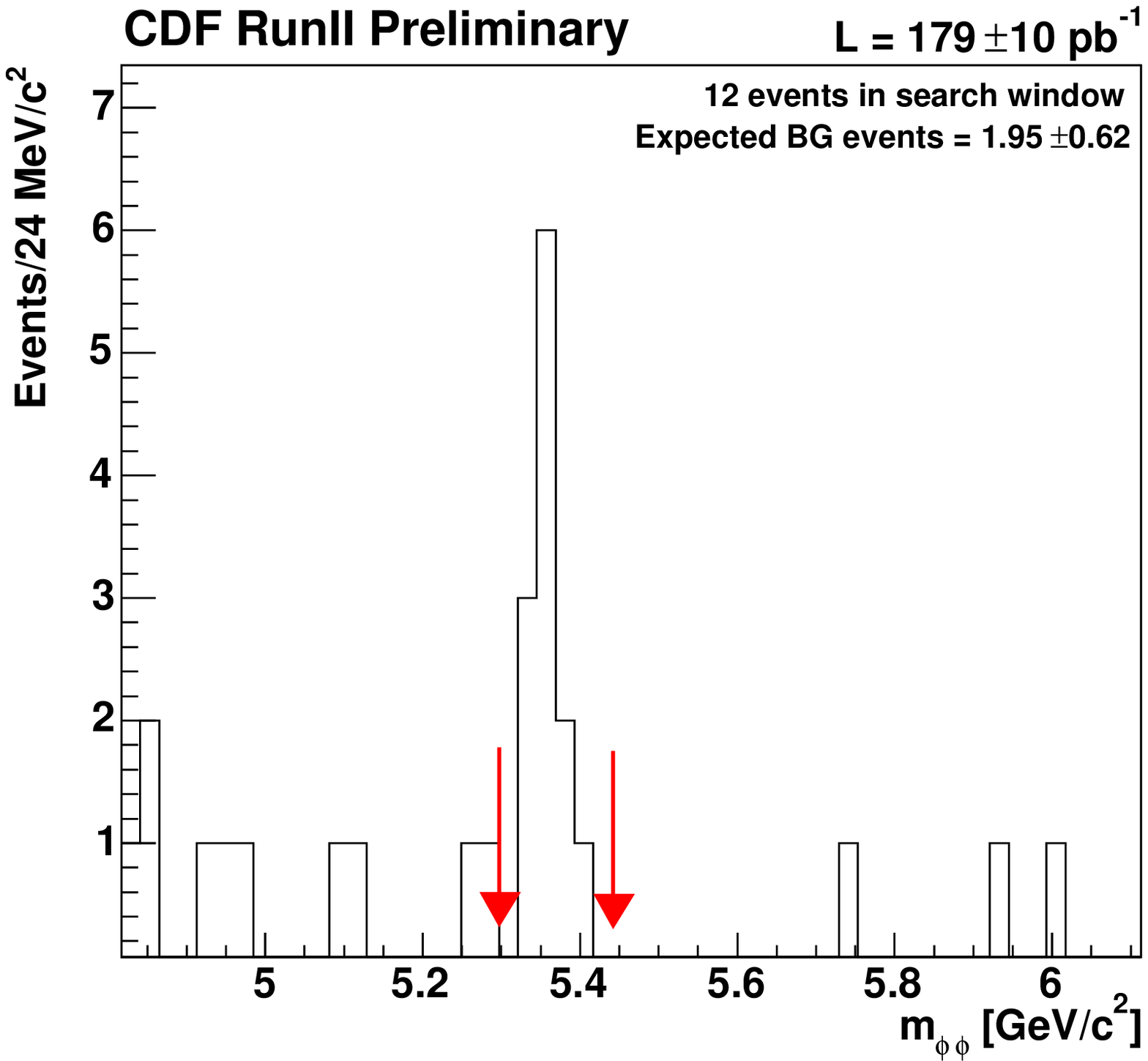}
\put(-280,105){\bf (a)}
\put(-120,105){\bf (b)}
}
\caption{\it
Invariant mass distribution of 
(a) $B^+\ra \phi K^+$ and (b) $\Bs\ra \phi\phi$.
}
\label{fig:penguin}
\end{figure}

CDF has shown examples of fully reconstructed hadronic $B$~decays from data
using the displaced track trigger (see e.g.~Ref.\cite{ref:mpprague}). We report
on a new search for charmless $B$~decays mediated by gluonic 
$b\ra s$~penguin decays. These decays are of interest in the light of
a possible contribution other than the usual mixing induced phase in the
time dependent $CP$~violation asymmetry observed at the $B$~factories. CDF
uses 180~pb$^{-1}$ of displaced track trigger data to search for 
$B^+\ra \phi K^+$ and $\Bs\ra \phi\phi$. Figure~\ref{fig:penguin}(a) shows the 
$\phi K^+$ invariant mass distribution indicating a signal of $(47.0\pm8.4)$
$B^+$ signal events. From this yield, CDF determines the ratio of branching
ratios 
${\cal B}(B^+\ra K^+\phi) / {\cal B}(B^+\ra J/\psi K^+)=(7.2\pm1.3\pm0.7)
\cdot 10^{-3}$ and the charge asymmetry 
${\cal A}_{CP}=-0.07\pm0.17^{+0.06}_{-0.05}$. Both results are 
in good agreement with the $B$~factories. 

The search for the never observed mode $\Bs\ra \phi\phi$ was performed in
a blind fashion using kinematically similar decays such as 
$B^0\ra J/\psi K^{*0}$ plus MC for cut optimization. 
Fig.~\ref{fig:penguin}(b) displays a signal of 12 events on an estimated
background of about 2 events. CDF determines 
${\cal B}(\Bs\ra\phi\phi)=(1.4\pm0.6\pm0.2\pm0.5_{BR})\cdot10^{-5}$ where the
error of $\pm0.5_{BR}$ results from the uncertainty in 
${\cal B}(\Bs\ra J/\psi\phi)$ used as normalization mode.

%-------------------------------------------------------------------
%--------------------------------------------------------------------

\subsection{Measurement of Hadronic Invariant Mass Moments}

\begin{figure}[tb]
\centerline{
\epsfysize=4.5cm
\epsffile{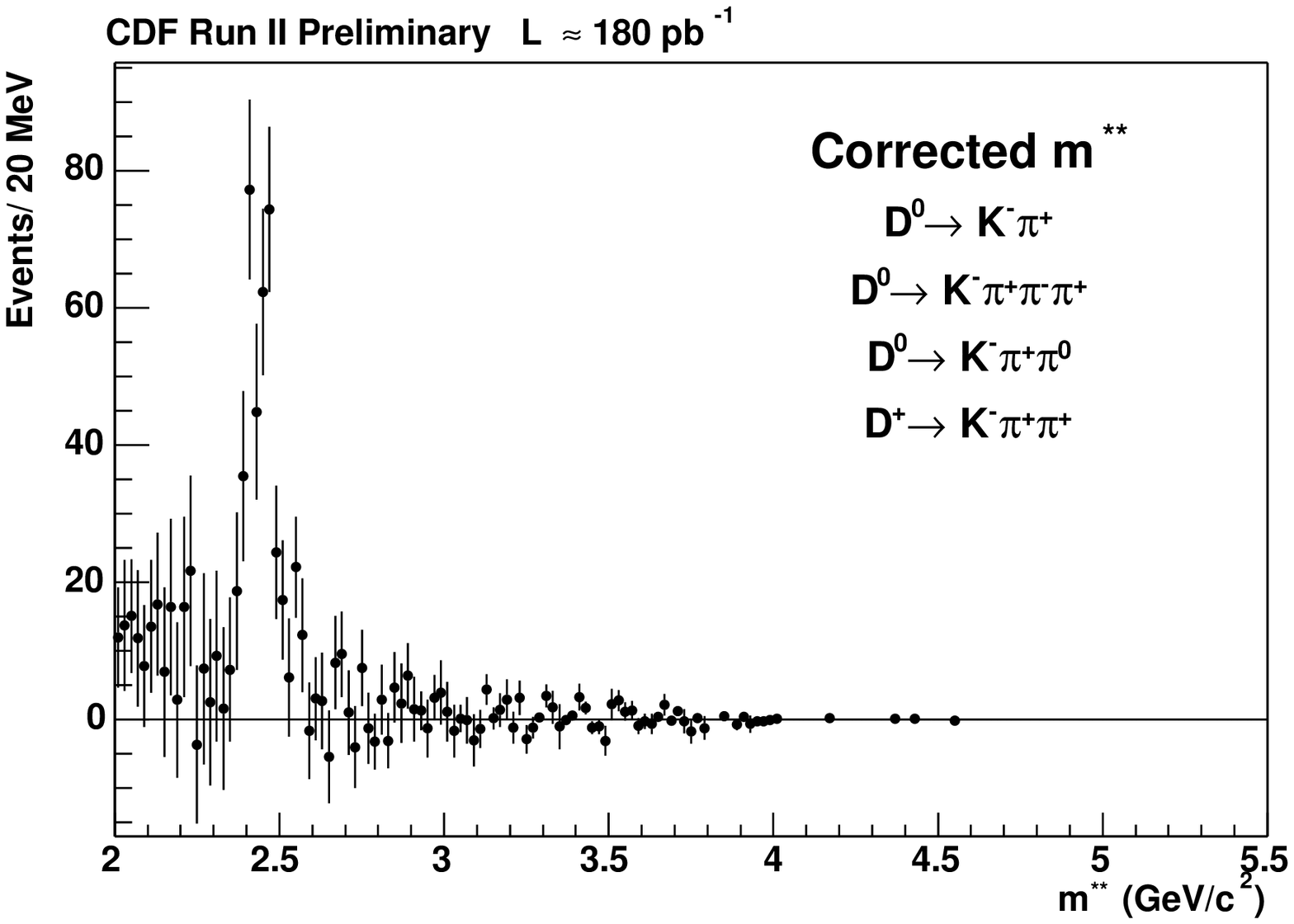}
\epsfysize=4.3cm
\epsffile{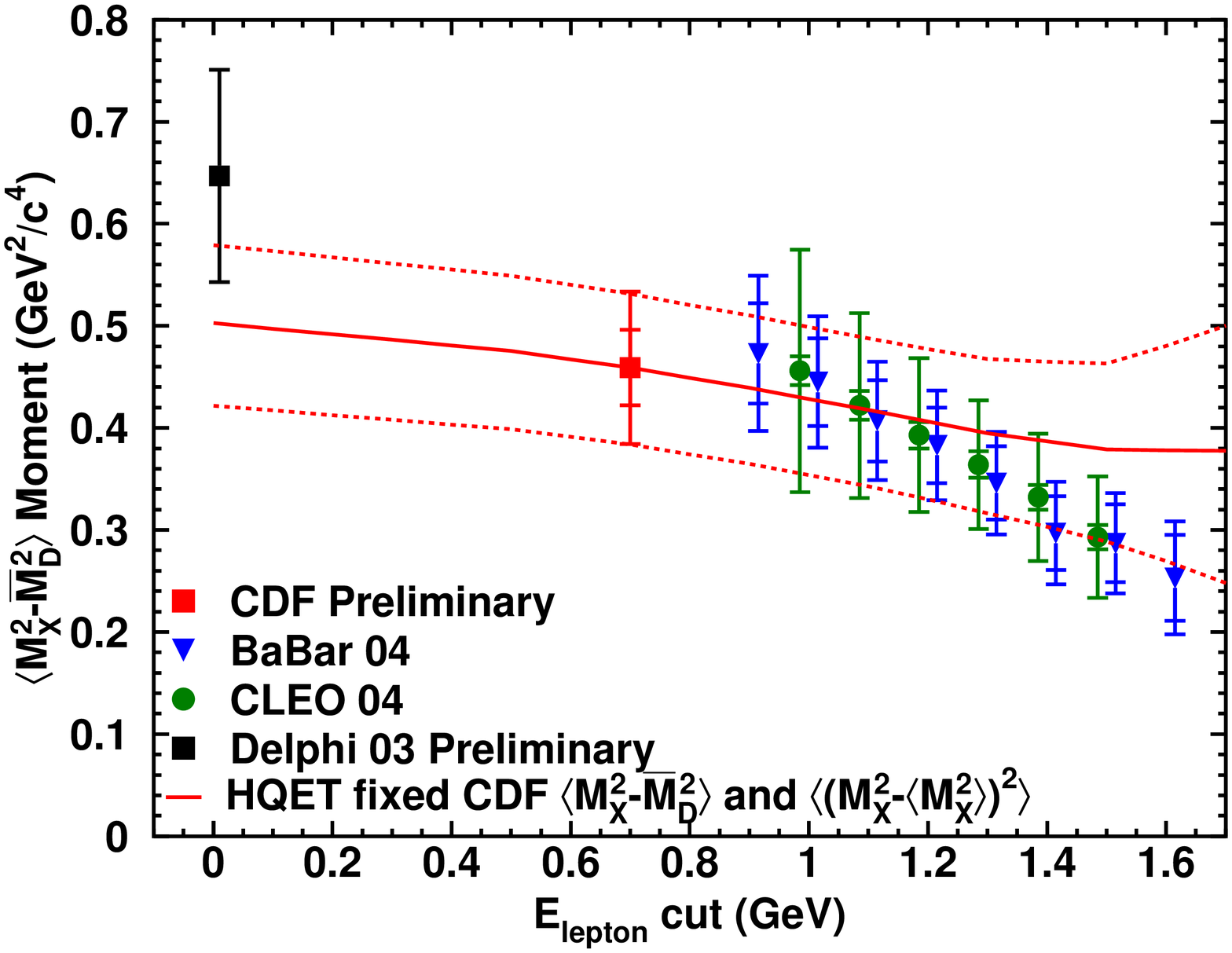}
\put(-290,100){\bf (a)}
\put(-30,100){\bf (b)}
}
\caption{\it
(a) Fully corrected invariant mass distribution $m(D^{(*)+}\pi^-_{**})$.
(b) Comparison between the CDF measurement of $M_1$ and previous
  determinations. 
}
\label{fig:mhadmom}
\end{figure}

Using 180~pb$^{-1}$ of data, CDF measured the first two moments of the
hadronic invariant mass squared distribution in semileptonic $B$~decays
using lepton plus SVT trigger data. Combining a direct measurement of the
$D^{**}$ piece -- see Fig.~\ref{fig:mhadmom}(a) for the fully corrected
inv.~mass distribution $m(D^{(*)+}\pi^-_{**})$ -- with the $D$ and $D^*$
pieces taken from PDG, CDF finds
$M_1\equiv\langle s_H\rangle-m^2_{\bar D}
=(0.459\pm0.037\pm0.019\pm0.062_{BR})$~GeV$^2$
and $M_2\equiv\langle(s_H-\langle s_H\rangle)^2\rangle 
=(1.04\pm0.25\pm0.07\pm0.10_{BR})$~GeV$^4$ where $0.062_{BR}$ and 
$0.10_{BR}$ refer to the
uncertainties coming from the branching ratios needed for the combination of
the $D$, $D^*$ and $D^{**}$ pieces.
Fig.~\ref{fig:mhadmom}(b) shows good agreement 
between the CDF measurement of $M_1$ and previous determinations.

%-------------------------------------------------------------------
%--------------------------------------------------------------------

\subsection{Observation of $X(3872)$}

%-------------------------------------------------------------------
\begin{figure}[btp]
\centerline{
\epsfysize=4.7cm
\epsffile{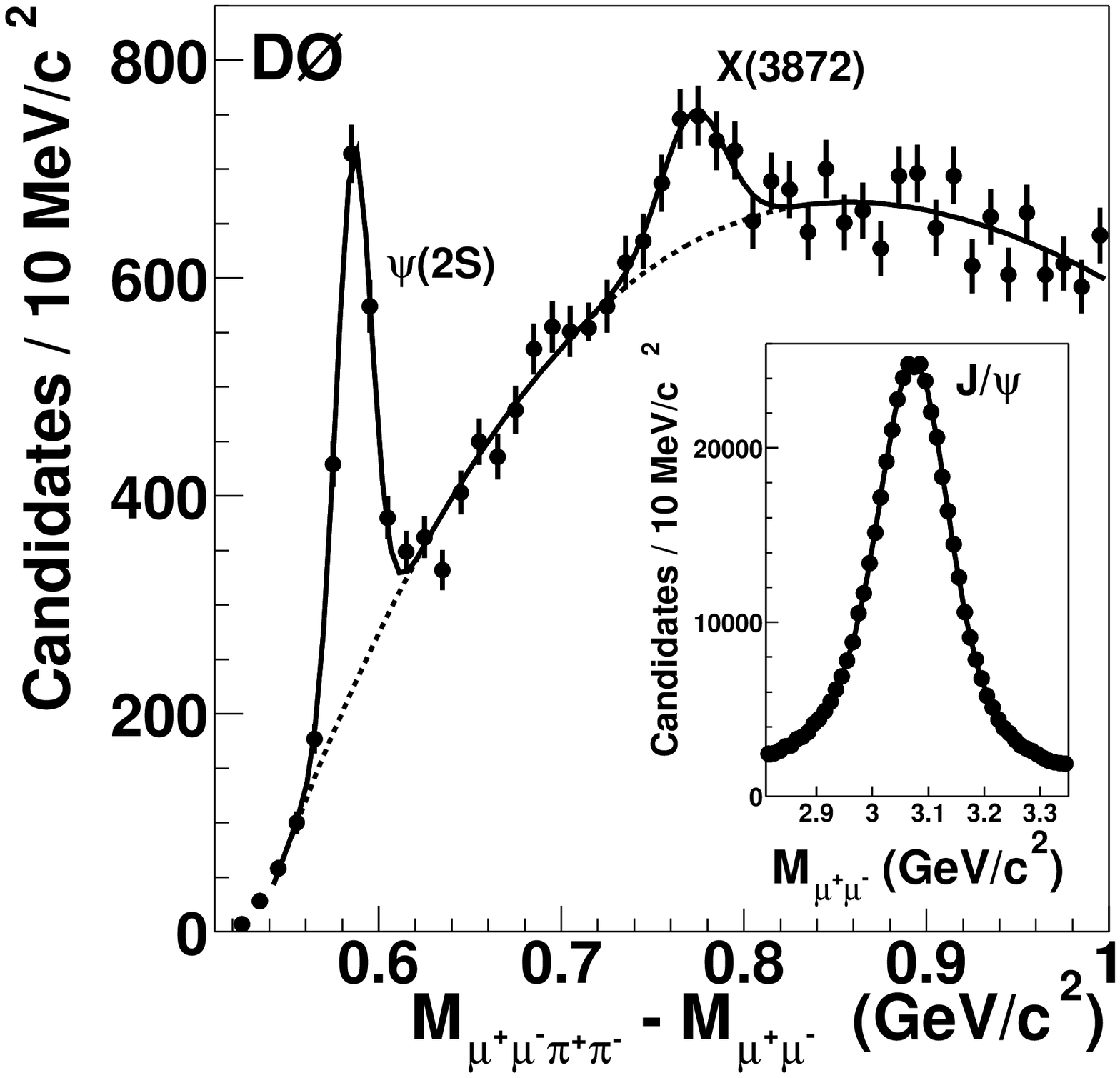}
\epsfysize=4.7cm
\epsffile{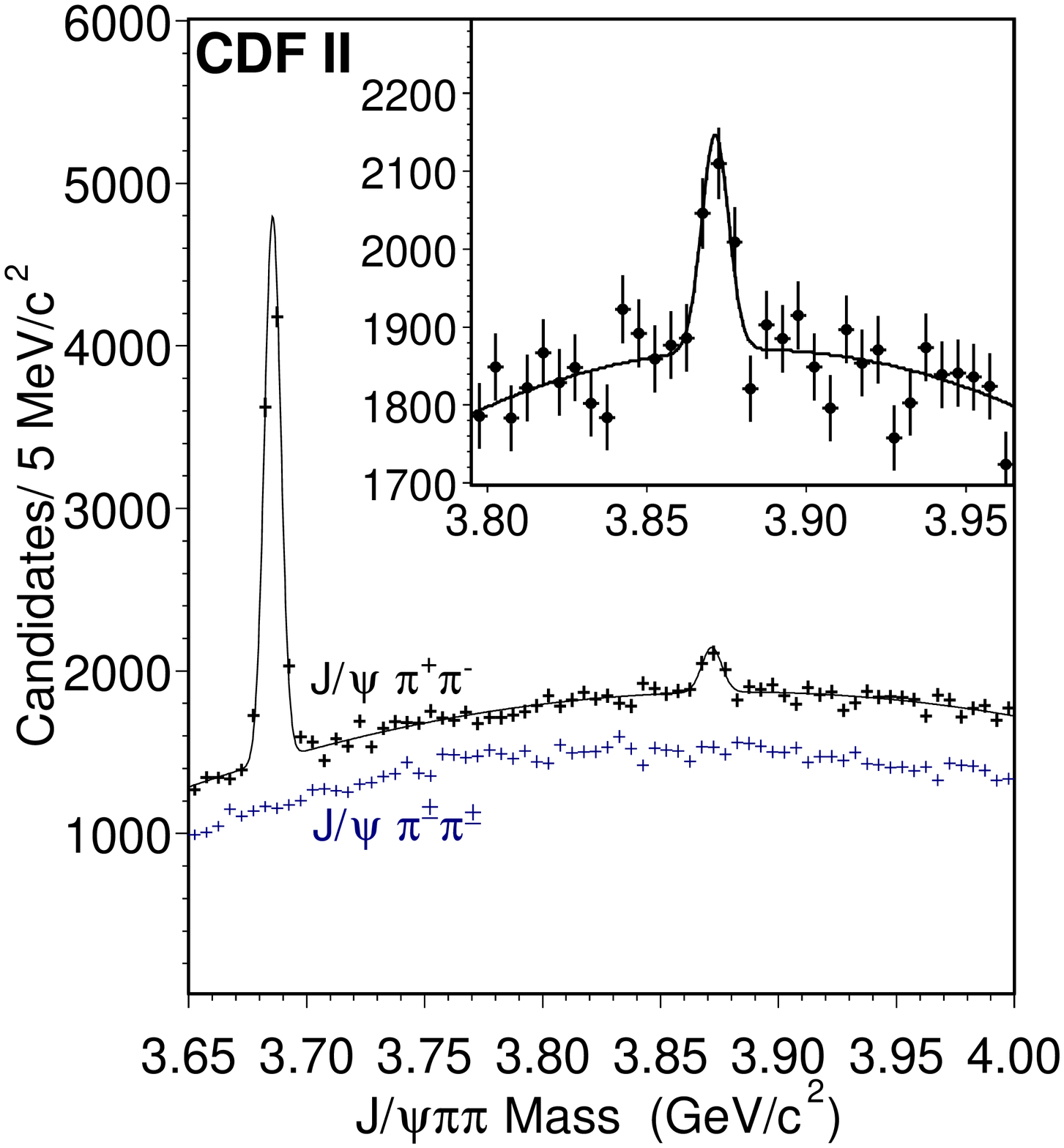}
\put(-215,28){\bf (a)}
\put(-30,28){\bf (b)}
}
\caption{\it
Mass distribution of $J/\psi\pi^+\pi^-$ candidates from 
(a) D\O\ and (b) CDF.  
}
\label{fig:xmass}
\end{figure}
%--------------------------------------------------------------------

Recently, the Belle collaboration reported a new particle $X(3872)$ 
observed\cite{ref:bellex}
in exclusive decays of $B$~mesons at a mass of 3872~\mevcc\ decaying into
$J/\psi\pi^+\pi^-$. 
The observation of this narrow resonance has been confirmed by the CDF
collaboration and recently also by the D\O\ experiment as shown in
Fig.~\ref{fig:xmass}. D\O\ observes $552\pm100\ X(3872)$ candidates and
measures the mass difference between the $X(3872)$ state and the $J/\psi$ to
be $(774.9\pm3.1\pm3.0)$~\mevcc. CDF observes $730\pm90$ events at a mass of 
$(3871\pm0.7\pm0.4)$~\mevcc\ with a width consistent with the detector
resolution. 

\subsection{Search for Pentaquark States}

\begin{figure}[tb]
\centerline{
\epsfxsize=4.0cm
\epsffile{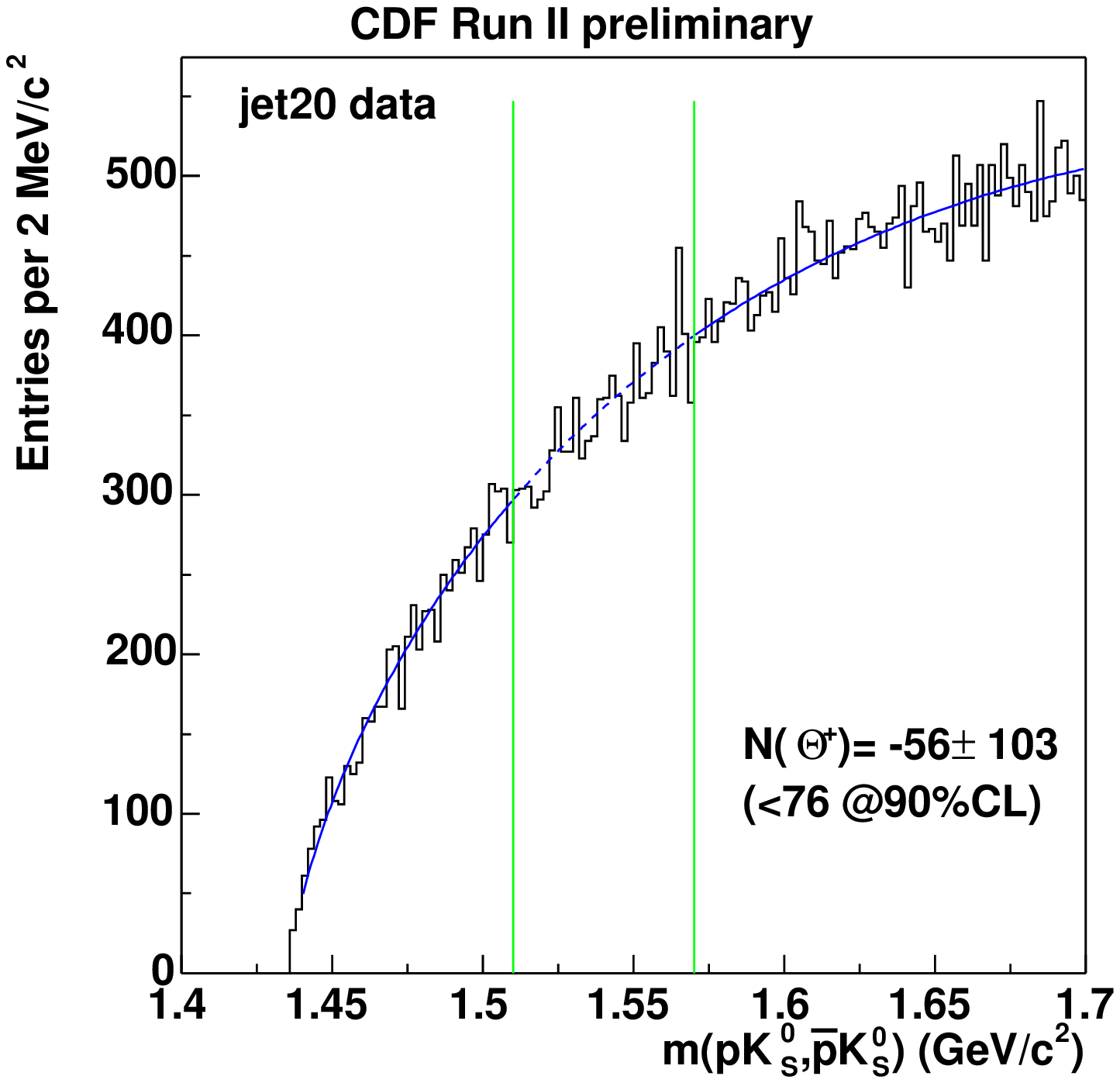}
\epsfxsize=4.0cm
\epsffile{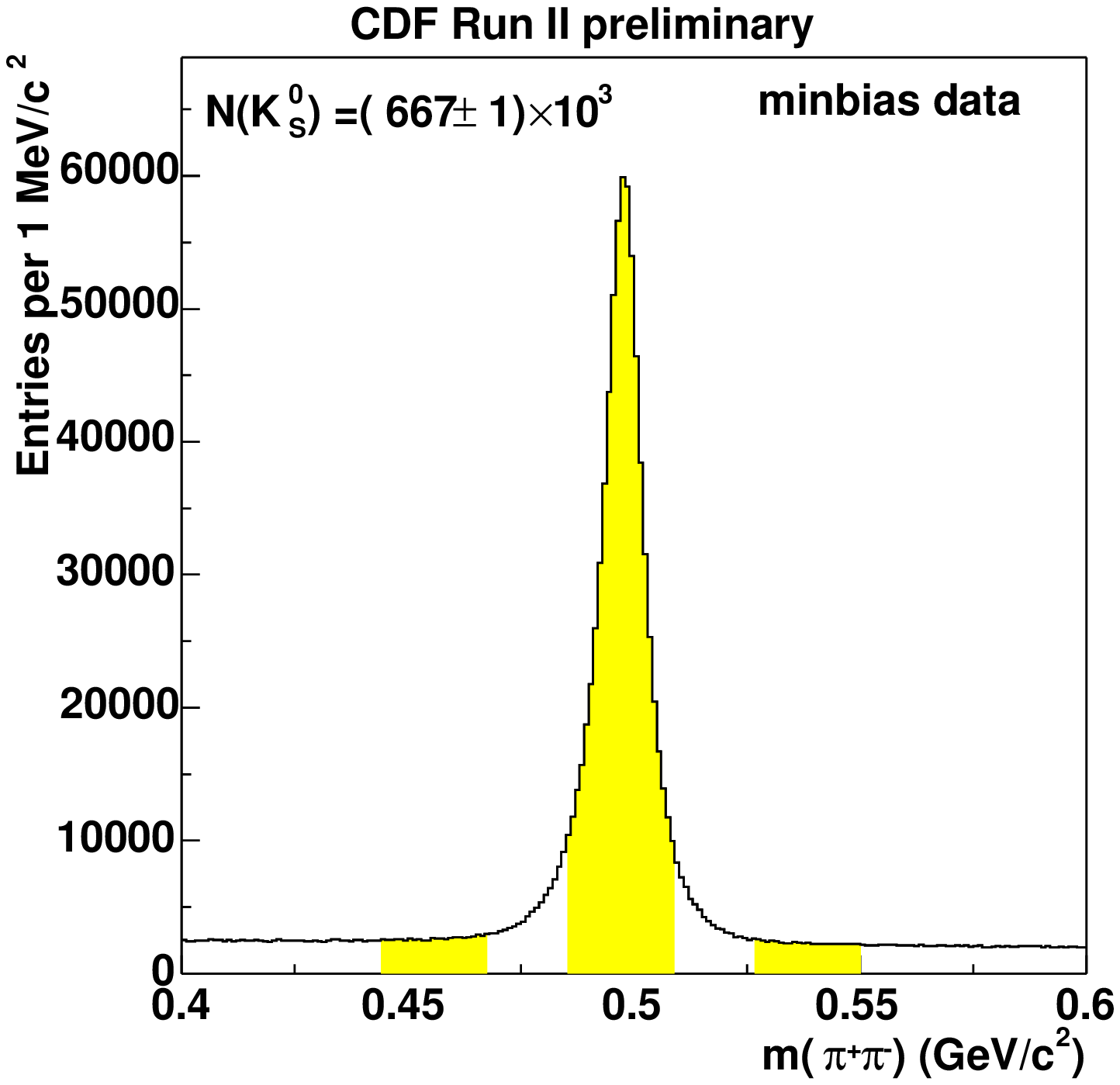}
\epsfxsize=4.0cm
\epsffile{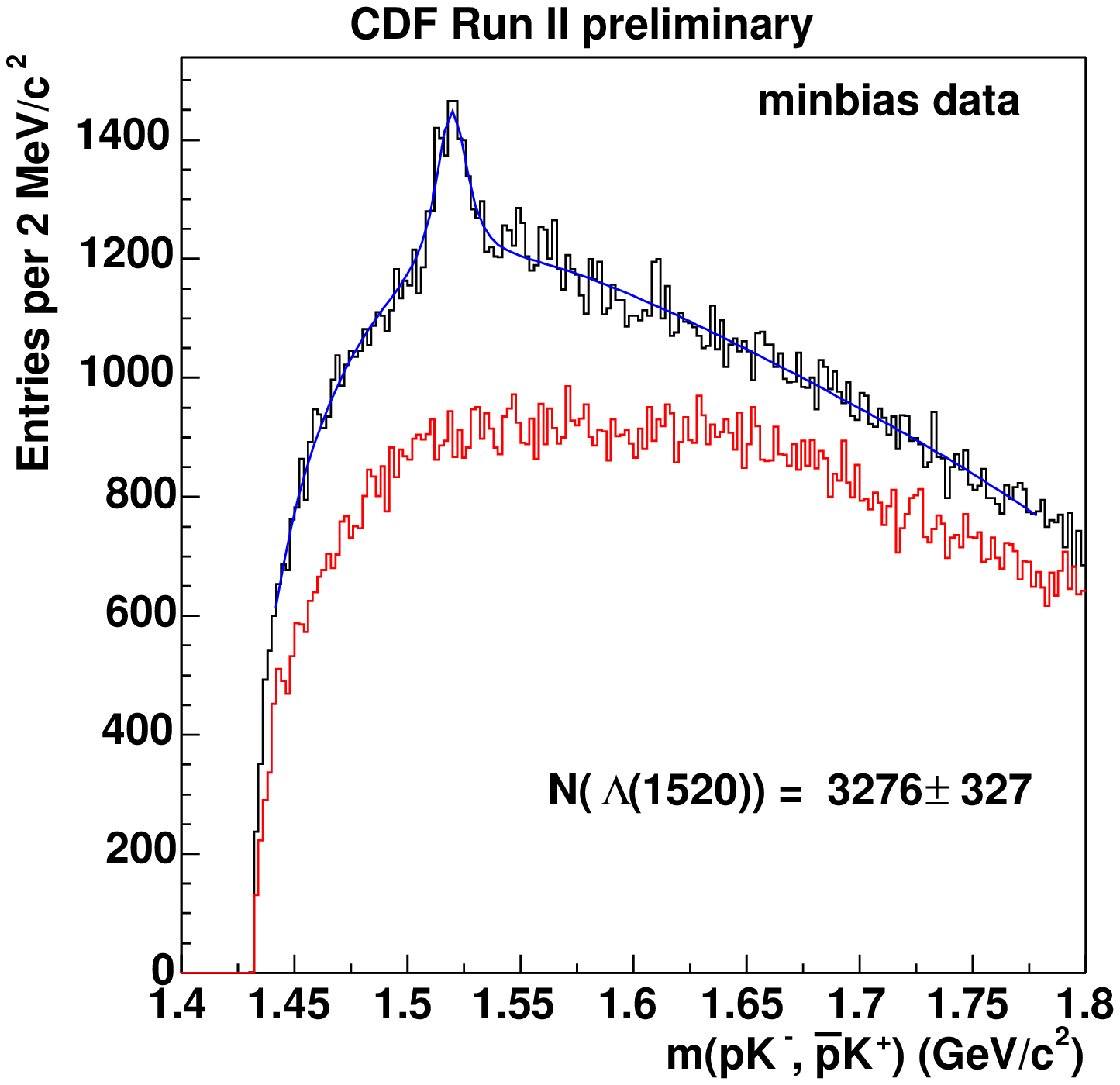}
\put(-324,83){\bf (a)}
\put(-145,83){\bf (b)}
\put(-25,83){\bf (c)}
}
\caption{\it
(a) Search for pentaquark state
$\Theta^+(1540)\ra p K_S^0$.
Reference states 
(b) $K^0_S\ra\pi\pi$ and 
(c) $\Lambda(1520)\ra pK^-$.
}
\label{fig:theta}
\end{figure}

\begin{figure}[tb]
\centerline{
\epsfxsize=6.0cm
\epsffile{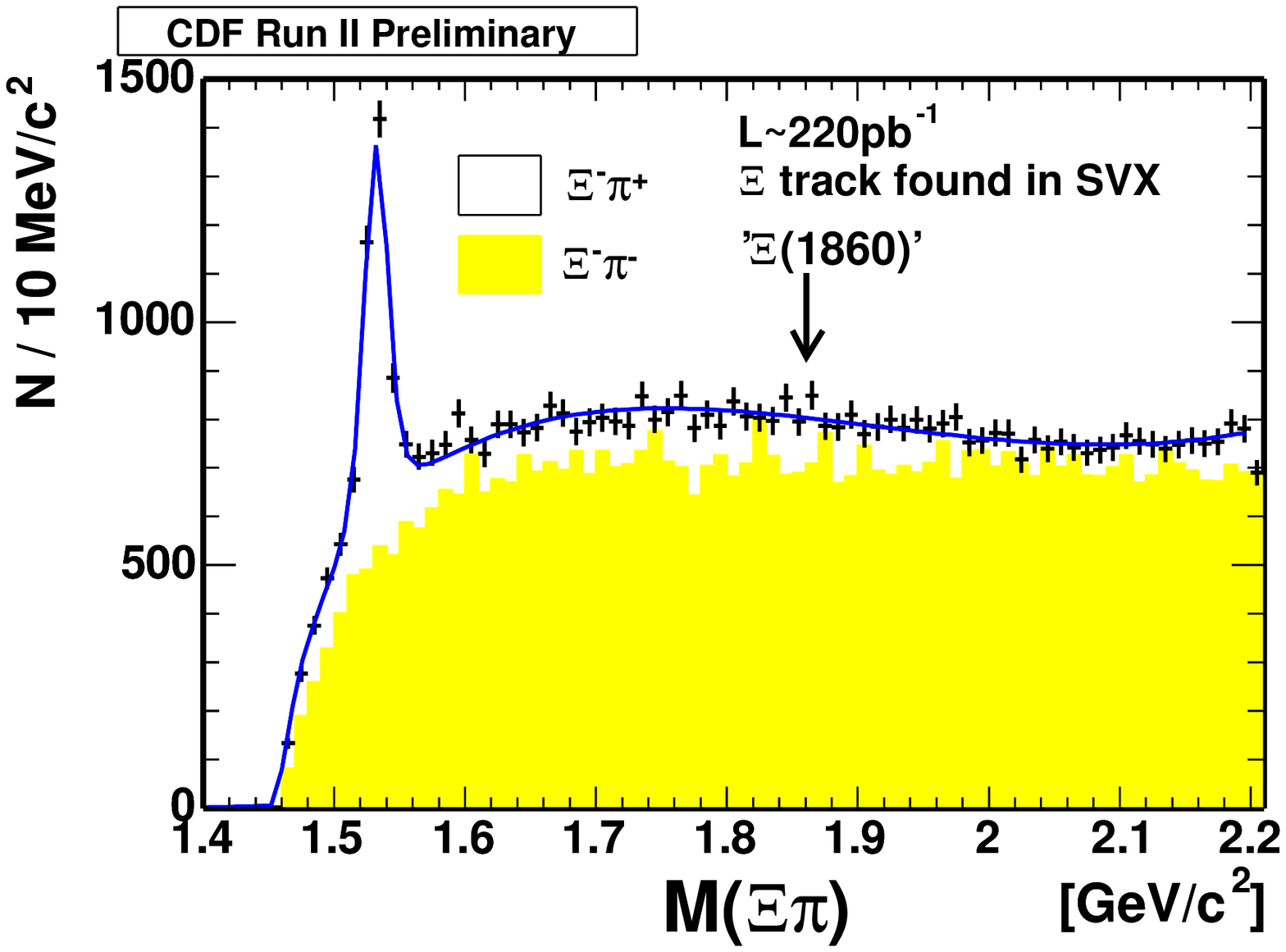}
\epsfxsize=6.0cm
\epsffile{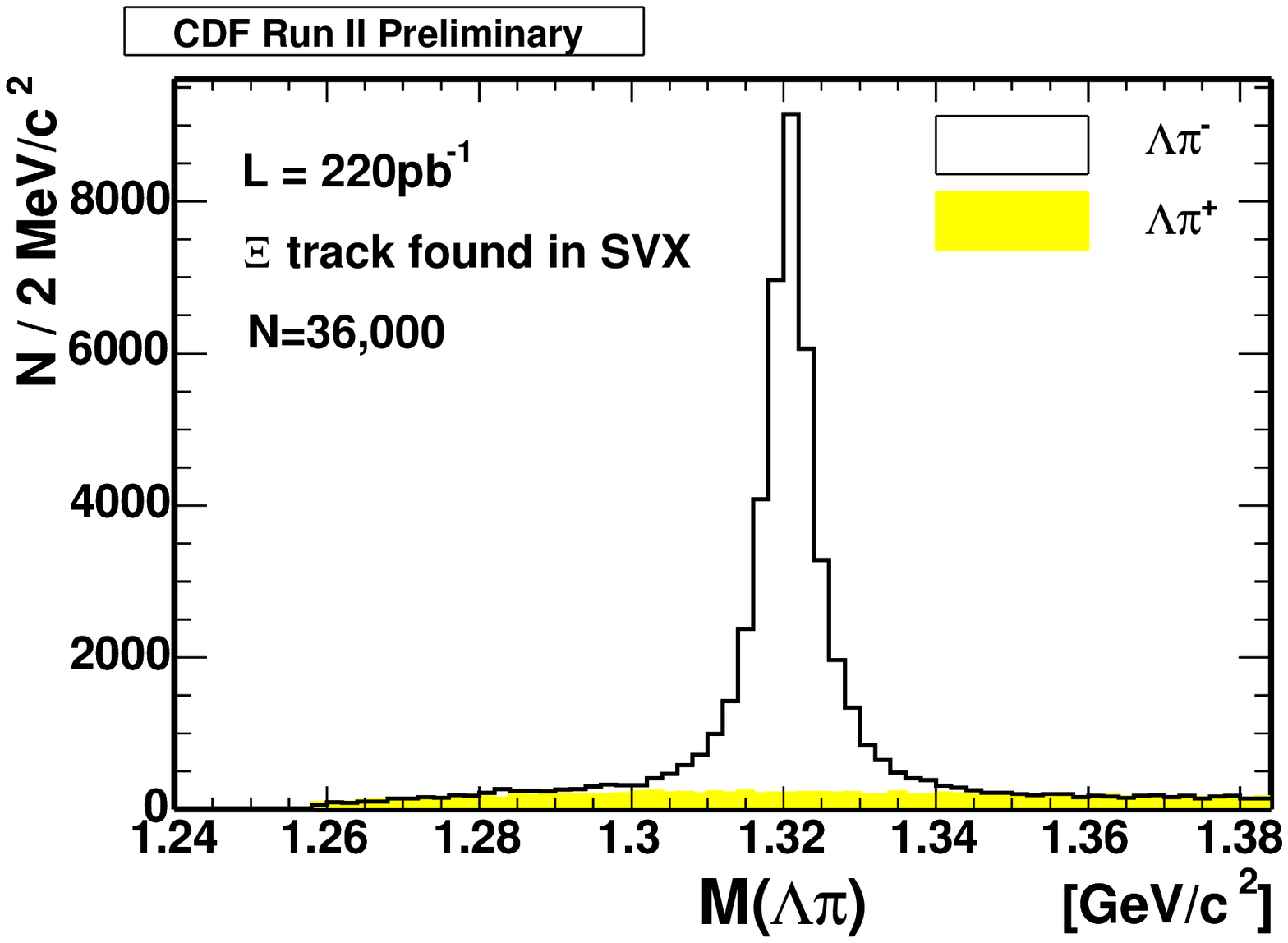}
\put(-202,85){\bf (a)}
\put(-30,80){\bf (b)}
}
\caption{\it
(a) Search for pentaquark state
$\Xi^{--}_{3/2}/\Xi^{0}_{3/2}\ra \Xi^- \pi^-/\pi^+$.
(b) Reference channel $\Xi\ra\Lambda\pi$.
}
\label{fig:xi}
\end{figure}

\begin{figure}[tb]
\centerline{
\epsfxsize=6.0cm
\epsffile{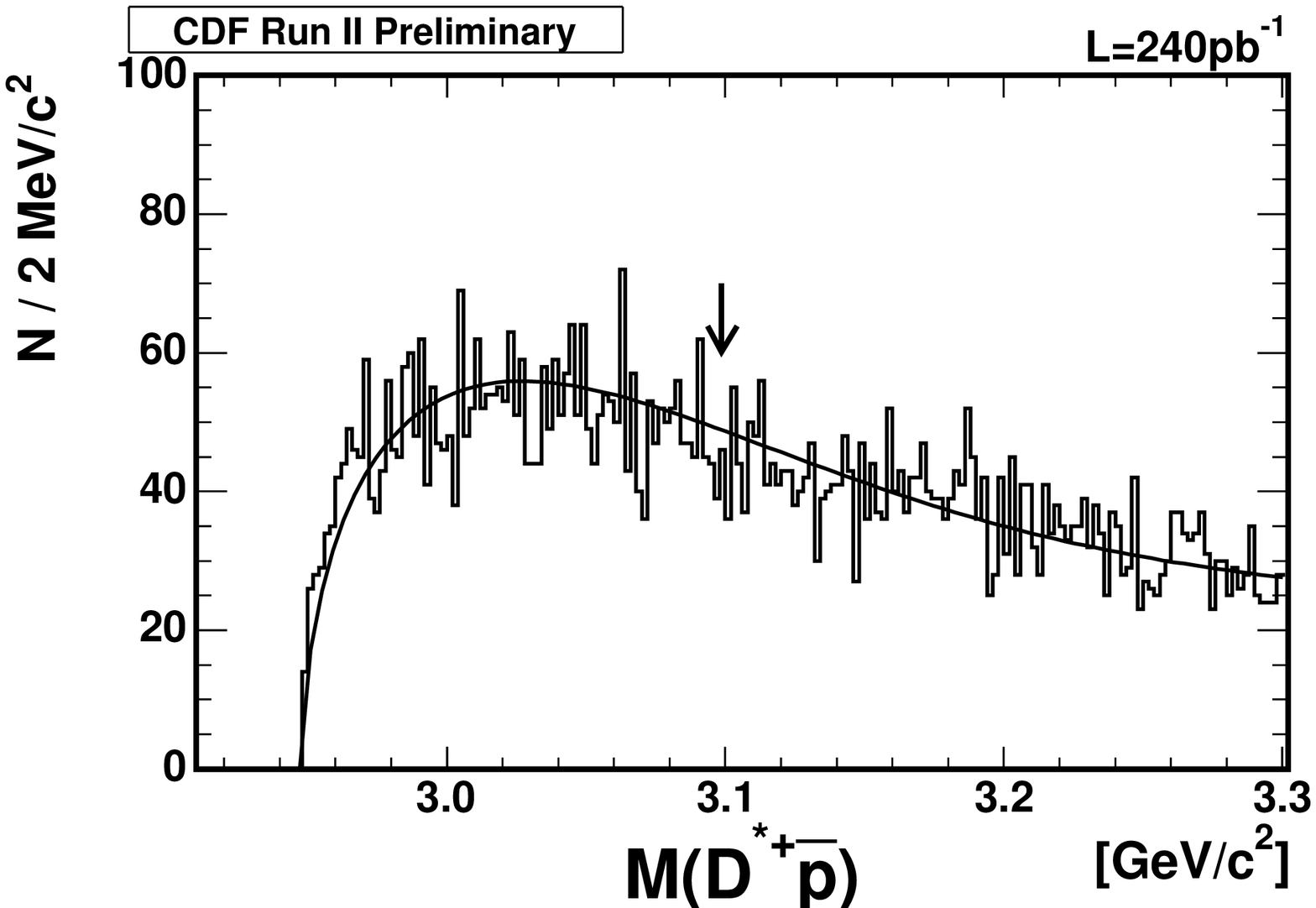}
\epsfxsize=6.0cm
\epsffile{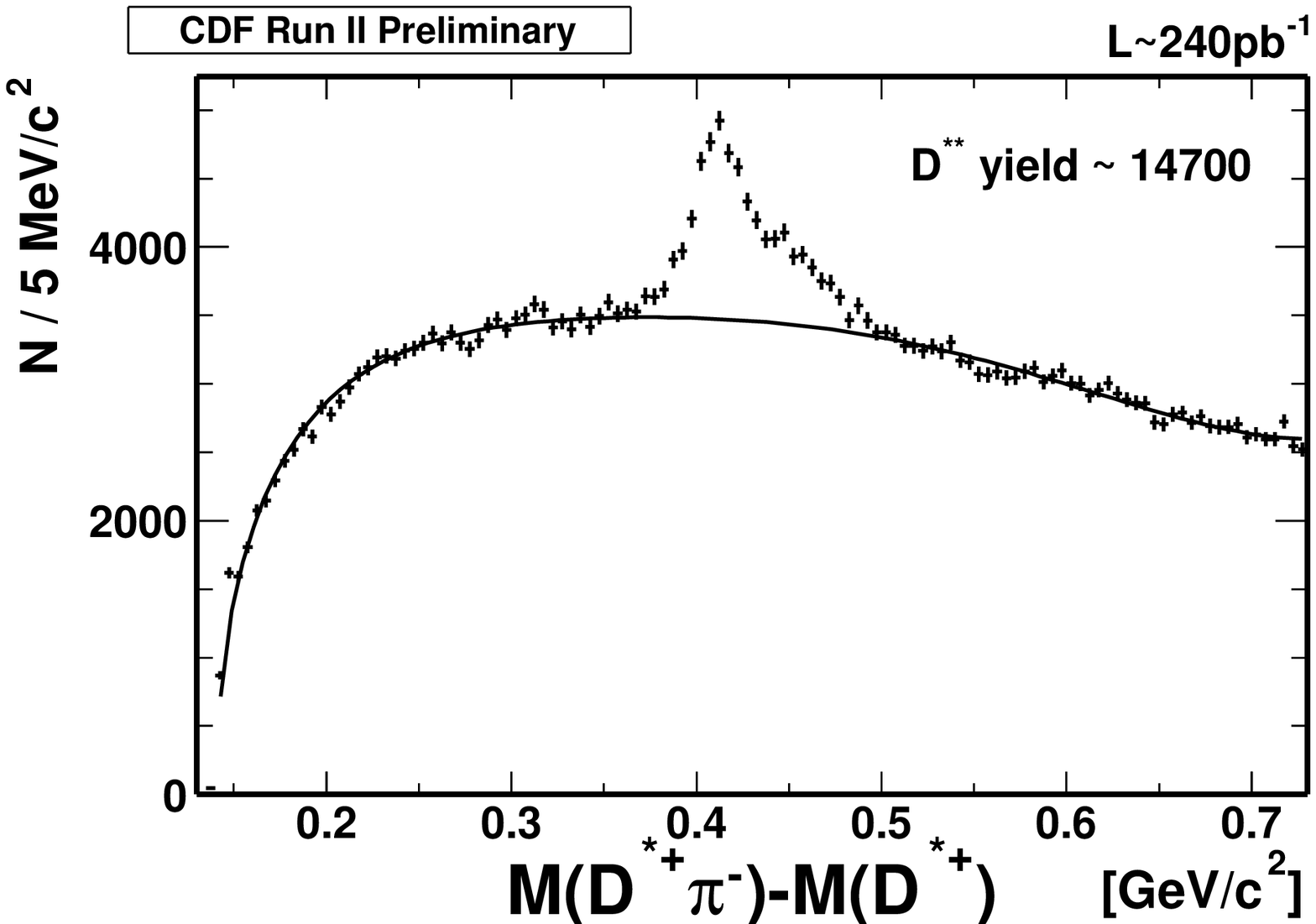}
\put(-310,92){\bf (a)}
\put(-135,92){\bf (b)}
}
\caption{\it
(a) Search for pentaquark state
$\Theta_c\ra p D^{*-}$.
(b) Reference channel $D^{*-}\pi^+$.
}
\label{fig:thetac}
\end{figure}

An exotic baryon, $\Theta^+(1540)$, with the quantum numbers of
$K^+n$ has recently been reported by several groups (for an overview, see
e.g.~Ref.\cite{ref:pentaquarks}). Such a state has a minimal quark content
of $|uudd\bar s\rangle$. Evidence for other pentaquark states such as 
an isospin 3/2 multiplet of $\Theta$'s with strangeness
$S=-2$\cite{ref:xi} and charmed pentaquarks\cite{ref:pentac} has also been
reported. CDF performed a search for the following pentaquark states:
$\Theta^+(1540)\ra p K_S^0$, 
$\Xi^{--}_{3/2}/\Xi^{0}_{3/2}\ra \Xi^- \pi^-/\pi^+$ and
$\Theta_c\ra p D^{*-}$. In each case a reference state has been
reconstructed. As shown in Figures~\ref{fig:theta}, \ref{fig:xi} and 
\ref{fig:thetac}, no evidence for a narrow signal has been found.
 
%-------------------------------------------------------------------
%--------------------------------------------------------------------

%-------------------------------------------------------------------
%--------------------------------------------------------------------

%-------------------------------------------------------------------
%--------------------------------------------------------------------

\section{Summary}

We review recent result on heavy quark physics at TeV energies focusing on
Run\,II measurements  at the Tevatron. A wealth of new $B$~physics
measurements from CDF and D\O\ has been reported. In particular, D\O\
demonstrates a very competitive $B$~physics program in Run\,II.

\section*{Acknowledgments}

I like to thank the organizers for a stimulating
meeting. It was a pleasure to attend this great conference.
I also would like to thank Ann, Emma and Helen, a constant source of
inspiration and support, for
their continuous understanding about the life of a physicist.


\begin{thebibliography}{99}

\bibitem{ref:myrevart} 
M.~Paulini,
Int.\ J.\ Mod.\ Phys.\ A {\bf 14} (1999) 2791
[hep-ex/9903002].

\bibitem{ref:docdfup} 
R.~Blair {\it et al.}  [CDF\,II~Collaboration],
%{\it ``The CDF-II detector: Technical design report,''}
FERMILAB-PUB-96-390-E (1996); \\
S.~Abachi {\it et al.} [D\O~Collaboration],
%{\it ``The D0 upgrade: The detector and its physics,''}
FERMILAB-PUB-96-357-E (1996).

\bibitem{ref:mpprague}
M.~Paulini, % (Representing the CDF and D0 Collaborations),
[hep-ex/0402020].

\bibitem{ref:bellex}
S.~K.~Choi {\it et al.}  [Belle Collaboration],
Phys.\ Rev.\ Lett.\  {\bf 91} (2003) 262001.

\bibitem{ref:pentaquarks}
D.~Diakonov, V.~Petrov and M.~V.~Polyakov,
Z.\ Phys.\ A {\bf 359} (1997) 305;\\
%[hep-ph/9703373]; \\
R.~L.~Jaffe and F.~Wilczek,
Phys.\ Rev.\ Lett.\  {\bf 91} (2003) 232003.
%[hep-ph/0307341].

\bibitem{ref:xi}
C.~Alt {\it et al.}  [NA49 Collaboration],
Phys.\ Rev.\ Lett.\  {\bf 92} (2004) 042003.
%[arXiv:hep-ex/0310014].

\bibitem{ref:pentac}
A.~Aktas {\it et al.}  [H1 Collaboration],
[hep-ex/0403017].

\end{thebibliography}
\end{document}